\documentclass[footinbib,a4paper,aps,pra,superscriptaddress,https://www.overleaf.com/project/5e6f523e2178db0001218358reprint,twocolumn,preprintnumbers,amsmath,amssymb,nobalancelastpage,10pt]{revtex4-1}
\usepackage{amsmath}
\usepackage{mathtools}
\usepackage{bbold}
\usepackage{verbatim}
\usepackage{graphicx}
\usepackage{color}
\usepackage{tikz}
\usepackage{subfigure}
\usepackage{float}
\usepackage{mathptmx}
\usepackage{array}
\usepackage{bm}
\usepackage{xcolor}
\usepackage{graphicx}
\usepackage{braket}
\usepackage[colorlinks=true,citecolor=blue,linkcolor=red, urlcolor=blue]{hyperref}

\usepackage{xcolor}
\usepackage[export]{adjustbox}

\usepackage{tikz,xcolor,hyperref}

\definecolor{lime}{HTML}{A6CE39}
\DeclareRobustCommand{\orcidicon}{
	\begin{tikzpicture}
	\draw[lime, fill=lime] (0,0) 
	circle [radius=0.16] 
	node[white] {{\fontfamily{qag}\selectfont \tiny ID}};
	\draw[white, fill=white] (-0.0625,0.095) 
	circle [radius=0.007];
	\end{tikzpicture}
	\hspace{-2mm}
}

\foreach \x in {A, ..., Z}{%
	\expandafter\xdef\csname orcid\x\endcsname{\noexpand\href{https://orcid.org/\csname orcidauthor\x\endcsname}{\noexpand\orcidicon}}
}

\begin{document}
	\title{Universal and optimal coin sequences for high entanglement generation in 1D discrete time quantum walks}
	
	\author{Aikaterini Gratsea\orcidA{}}\email{gratsea.katerina@gmail.com}
	\affiliation{Quantum Systems Unit, Okinawa Institute of Science and Technology Graduate University, 1919-1 Tancha, Onna, Okinawa 904-0495, Japan}
	\affiliation{ICFO - Institut de Ciencies Fotoniques, The Barcelona Institute of Science and Technology, Av.~Carl Friedrich Gauss 3, 08860 Castelldefels (Barcelona), Spain}
	
	\author{Friederike Metz\orcidB{}}\email{friederike.metz@oist.jp}
	\affiliation{Quantum Systems Unit, Okinawa Institute of Science and Technology Graduate University, 1919-1 Tancha, Onna, Okinawa 904-0495, Japan}
	
	\author{Thomas Busch\orcidC{}}
	\affiliation{Quantum Systems Unit, Okinawa Institute of Science and Technology Graduate University, 1919-1 Tancha, Onna, Okinawa 904-0495, Japan}
	
	\begin{abstract}
		
		Entanglement is a key resource in many quantum information applications and achieving high values independently of the initial conditions is an important task. Here we address the problem of generating highly entangled states in a discrete time quantum walk irrespective of the initial state using two different approaches. First, we present and analyze a deterministic sequence of coin operators which produces high values of entanglement in a universal manner for a class of localized initial states. In a second approach, we optimize the discrete sequence of coin operators using a reinforcement learning algorithm. While the amount of entanglement produced by the deterministic sequence is fully independent of the initial states considered, the optimized sequences achieve in general higher average values of entanglement that do however depend on the initial state parameters. Our proposed sequence and optimization algorithm are especially useful in cases where the initial state is not fully known or entanglement has to be generated in a universal manner for a range of initial states.
		
	\end{abstract}
	
	\maketitle

\section{\textbf{Introduction}} 

Entanglement plays a fundamental role in quantum information processing \cite{Horodecki2009} and maximising it is an important goal for reaching high fidelity operations. Developing and understanding methods for creating, maximising and in general engineering entanglement are therefore important topics in research at the moment. While most often entanglement between different states of the same degree of freedom is considered, more recently so-called hybrid entanglement between different degrees of freedom has been of interest \cite{Sciarrino2018}. If these degrees of freedom belong to the same particle, more information can be encoded at the single particle level, which can help to reduce required resources. Hybrid entanglement has recently been experimentally created in photonic architectures \cite{Zeilinger2018, Neves_2009} and also studied in neutrons \cite{Rauch2007} and bosonic atoms \cite{Smerzi2018}.

Here, we discuss the generation of hybrid entangled states that can be created in discrete time quantum walks, where the motion of a particle that moves in a high-dimensional discrete space depends on an internal, two-dimensional coin degree of freedom \cite{Aharonov1993,Kempe2003,Venegas2012}. The evolution itself consists of the recurrent application of a coin and shift operator which in general leads to entanglement between the walker and coin. Quantum walks have already been realized in a variety of physical systems such as cold atoms \cite{Preiss2015, Mugel2016}, trapped ions \cite{Schmitz2009,Roos2010}, superconducting qubits \cite{Zhou2019, Siddigi2017}, neutral atoms \cite{Karski2009, Robens_2016}, nuclear magnetic resonance systems \cite{Laflamme2005, Rongdian2003} and photonic architectures \cite{Neves_2018, Sciarrino2019, Crespi2013}. Hybrid entanglement generation has been observed as well \cite{Cardano2015, Guo2018}, and has been used as a resource for quantum teleportation \cite{Chatterjee2019} and for the design of secure communication protocols \cite{Srikara2020}.

Recently different approaches to enhance the entanglement between the walker and coin have been explored. It was shown that disorder in the coin can increase the amount of hybrid entanglement created by the walk \cite{Rigolin2013,Rigolin2014,Zeng2017,Chandrashekar2013}, and that randomly choosing the coin operator at each time step of the quantum walk can lead to maximally entangled states in the asymptotic limit independent of the initial state \cite{Rigolin2013}. However, the large number of steps this strategy requires makes the scheme unrealistic for current experiments. As a possible solution, the optimization of the coin operator sequence was suggested and it was shown that this can reduce the number of steps to less than 10 \cite{Innocenti2017,Gratsea2019}. However, the entanglement that can be generated by optimising is highly dependent on the initial state and requires potentially the full set of possible coin operators to be realized experimentally. In an alternative approach, Wang \textit{et al.}~suggested to restrict the set of possible coin operators to just the Hadamard and Fourier coins and showed that certain sequences give rise to highly entangled states with as few as 20 steps \cite{Guo2018}. However, the optimal sequences they found were also  highly dependent on the initial state.

In this work we present and discuss deterministic coin sequences that allow to create large amounts of hybrid entanglement in a quantum walk. To be experimentally realistic we restrict ourselves to Hadamard  and Fourier coins only and aim at a minimal number of steps. It is worth noting that deterministic sequences of coin operators have already been studied in the context of the localization-delocalization transition in quantum walks \cite{Kumar2018,Gullo2017, Ribeiro2004}, where the considered sequences range from the periodic cases \cite{Kumar2018} to aperiodic ones like the Thue-Morse, Rudin-Shapiro, and the Fibonacci sequence \cite{Gullo2017, Ribeiro2004}. 

The first sequence we discuss is designed to create the same, large amount of entanglement independently of the localized initial states, as long as they have a vanishing relative phase. The structure of the sequence allows it to work for any odd number of time steps, and it therefore also fulfils the requirement to be useful in experimental settings where often only a few steps can be realised. We further show that the amount of achievable entanglement can be controlled by replacing the Hadamard coin in the sequence by a general rotation operator. For localized initial states with nonzero local phases the same universal entangling behavior can be observed if the coin operators in the sequence are modified slightly. 

In the second part of this work, we ask if higher values of hybrid entanglement can be achieved through a direct optimization of the coin operator sequence and choose a reinforcement learning (RL) based approach for the optimization. Machine learning has already achieved remarkable results in various areas of physics \cite{Carleo2019,Briegel2017}, and different machine learning approaches have been combined with quantum walks for exploring quantum speed-up \cite{Melnikov2019,Briegel2014} and graph structures \cite{Dernbach2019}. On the other hand, RL has been successfully applied to challenging problems in quantum physics including quantum state preparation \cite{Bukov2018, Bukov2018b}, quantum optimal control \cite{Wang2019, Niu2019}, and quantum error correction \cite{Sweke2018, Foesel2018}.
 
In this work, we use a RL technique to tackle the optimization of entanglement in quantum walks. In contrast to previously employed optimization schemes, RL allows us to not only find the optimal sequence of coin operators for a specific initial state but also for classes of initial states.   The resulting optimized coin sequences achieve equal or higher average values of entanglement than the deterministic sequence discussed in the first part of this work. However, the amount entanglement created this way is not independent of the initial state.

The paper is organized as follows. In Sec.~\ref{Sec:Background} we briefly review the discrete time quantum walk and introduce the reinforcement learning framework as well as the Q learning algorithm used for optimization. In Sec.~\ref{Sec:Results} we present and analyze the universal entangling coin sequence and compare it to the results of the RL optimization problem. Finally, we conclude with a summary and outlook in Sec.~\ref{Sec:Conclusions}.

\section{Theoretical background}
\label{Sec:Background}

\subsection{\textbf{The quantum walk}}

The discrete time quantum walk in one dimension is realized on the tensor product of two Hilbert spaces $H = H_w \otimes H_c$ \cite{Aharonov1993,Kempe2003,Venegas2012}. The space corresponding to the position of the walker $H_w$ is high-dimensional and spanned by $\{\ket{x}: x \in \mathbb{Z}\}$, while the coin space $H_c$ is two-dimensional and spanned by $\{\ket{\uparrow}, \ket{\downarrow}\}$. We assume that the walker is initially localised on one site in an arbitrary superposition of the coin states
\begin{equation}
\vert \psi_0 \rangle =  \cos(\theta/2) \ket{0,\uparrow} + e^{i\phi}\sin(\theta/2) \ket{0,\downarrow}
\label{initial_general},
\end{equation}
where $\theta\in [0,\pi]$ and $\phi \in [0,2\pi]$. The evolution consists of $n$ applications of a unitary operator $U = S C$, where $S$ is a translation and $C$ is a local rotation. The translation $S$ moves the walker either to the left or to the right depending on the internal coin state and has the form
\begin{equation} \label{uniS}
 S = \sum_x  \ket{x-1,\uparrow}\bra{x,\downarrow} + \ket{x+1,\downarrow} \bra{x,\uparrow}.
\end{equation}
The coin operator $C$ rotates the inner degree of freedom and in its most general form can be expressed as

\begin{equation}\label{coin}
C =\mathbb{1}_x \otimes e^{i\beta} \left[\begin{array}{rr}
e^{i\xi}\cos(\alpha) & e^{i\zeta}\sin(\alpha) \\
-e^{-i\zeta}\sin(\alpha) & e^{-i\xi}\cos(\alpha)
\end{array}\right],
\end{equation}
where $\xi,\zeta \in [0, 2 \pi]$ and $ \alpha \in [0, \pi/2]$ are the parameters of the SU(2) rotation and $ \beta $ fixes the global phase~\cite{Chandrashekar2008}. The coin operators we want to employ are the Hadamard coin $H$
$[\beta=\pi/2 , \alpha=\pi/4 , \xi=\zeta=-\pi/2]$ and the Fourier coin $F$ $[\beta=0, \alpha=\pi/4 , \xi=0 , \zeta=\pi/2]$, which have the explicit forms
\begin{align} 
\label{Eq:Coins}
H =\frac{1}{\sqrt{2}}\left[\begin{array}{rr}
1 & 1 \\
1 & -1
\end{array}\right] ,
\qquad
F=\frac{1}{\sqrt{2}}\begin{bmatrix}
1 & i \\
i & 1
\end{bmatrix} .
\end{align}  

\subsection{Reinforcement learning} \label{Sec:RL}

Reinforcement learning (RL) is a sub field of machine learning in which a trainable agent interacts with an environment, takes actions, observes states, and obtains rewards \cite{Sutton1998}. The objective for the RL agent is to choose actions at each time step that maximize the expected future reward. Here, we implement the off-policy Q-learning algorithm with the goal of maximizing the hybrid entanglement between the walker and the coin degree of freedom \cite{Watkins1992}.

Each training episode consists of a fixed number of time steps $n$ of the quantum walk. Since we are interested in maximizing the entanglement after the evolution is complete, we set all rewards at intermediate time steps to zero and allow for a nonzero reward only at the final time step. We use the Schmidt norm as a measure of entanglement and therefore the reward $R$ at the end of each episode $i$ can be defined as
\begin{equation}
    R_i=  \sum_{k=1}^{K} \lambda_k.
\end{equation}
Here $\lambda_k$ are the Schmidt coefficients and $K=\text{min}(d_c,d_w)$ with $d_c$ and $d_w$ being the dimensions of the coin and walker subsystems respectively. Since the coin space is always two dimensional, we have $K=2$ and the rewards take values between $\it{R_i} \in [1, \sqrt{2}]$.

At each time step of the quantum walk, the agent can choose between two actions defined as $\it{A} \in \{H,F\}$, where $H$ and $F$ correspond to the Hadamard and Fourier coin operator respectively (see Eq.~\eqref{Eq:Coins}). This choice is made after obtaining information about the current state of the environment. One way of defining the RL state would be to use the full quantum state of the system at each time step of the quantum walk. However, since the quantum state is essentially a vector of continuous complex numbers, it cannot be straightforwardly employed in tabular (discrete) RL settings and more sophisticated methods like neural network function approximators are needed \cite{Sutton1998}. However, in our case we can use the fact that the dynamics of the system are deterministic and therefore the history of actions (applied coins) contains the same information for a fixed initial state. Specifically, for a given number of time steps $n$ and a specific initial state $\psi_0$, there are $2^n$ possible sequences. For example, for the case $n=2$ the complete set of sequences are $\{HH\psi_0, HF\psi_0, FH\psi_0, FF\psi_0\}$. Hence, no information about the intermediate physical states is needed and the RL states are simply given by $S \in \{\text{init},H, F, HH, HF, FH, FF\}$. Here, init refers to the initial state of the environment before the quantum walk evolution has started.

For obtaining the optimal policy $\pi^*(S) = A$, which indicates the optimal action to take given the current state, we employ the Q-Learning algorithm which is based on learning an optimal Q function. The Q value $Q^{\pi}(S,A)$ of a state-action pair is defined as the expected cumulative future reward when starting in state $S$, taking action $A$, and following the policy $\pi$ thereafter
\begin{equation}
Q^{\pi}(S,A) \doteq \mathbb{E}_{\pi}\left[\sum_{i}^{}  R_{i} \Big| S, A\right].
\end{equation}
Therefore, the Q value $Q(S,A)$ is a measure of how promising it is to choose the respective action $A$ in a state $S$. The optimal Q value $Q^*(S,A)$ is simply defined as the maximum Q value over all policies $Q^{*}(S, A)=\max _{\pi} Q^{\pi}(S,A)$. In case the optimal Q function is known for all state-action pairs, the optimal policy can be inferred by selecting actions that maximize the Q value, i.e.~a greedy action selection
\begin{equation}\label{Eq:greedy}
\pi^*(S)=\arg \max _{A} Q^{*}(S, A).
\end{equation}
Hence, it suffices to learn the optimal Q values which can be achieved through an iterative update rule known as Temporal Difference learning
\begin{equation}
    Q(S_i,A_i) \rightarrow Q(S_i,A_i) + \alpha \left[ R_i + \max_{A} Q(S_{i+1}, A) - Q(S_i,A_i) \right],
\end{equation}
where $\alpha \in [0,1]$ is the learning rate and the term in the brackets is called the target. It can be shown that the Q values eventually converge to their optimal values if the policy that is followed during training has a finite probability of visiting all state-action pairs \cite{Sutton1998}. Here, we use an $\epsilon$-greedy action selection during training, i.e.~the agent acts randomly with probability $\epsilon$ and otherwise takes action $A_i$ which maximizes the Q value in the current state: $A_i = \text{argmax}_A Q(S_i,A)$. Moreover, for a better trade-off between exploration of the full action space and exploitation of high rewards, $\epsilon$ is exponentially decaying after each training episode~$i$
\begin{equation}
    \epsilon(i) = (\epsilon_{\text{init}} - \epsilon_{\text{fin}})  \exp\left[\frac{-8 i}{N_{\text{episodes}}}\right] + \epsilon_{\text{fin}},
\end{equation}
with $\epsilon_{\text{init}}$ and $\epsilon_{\text{fin}}$ being the initial and final value of $\epsilon$, respectively. The exponential decay ensures that at the beginning of training the agent acts mostly random and explores a variety of different actions while towards the end of training actions are chosen more deterministically according to the target policy. Once training has successfully converged, the optimal policy is given by a fully greedy action selection given through Eq.~\eqref{Eq:greedy}.

\section{Hybrid Entanglement Creation}
\label{Sec:Results}

\subsection{Universal entangling coin sequence}

\begin{figure}[t!]
	\includegraphics[scale=0.52, left]{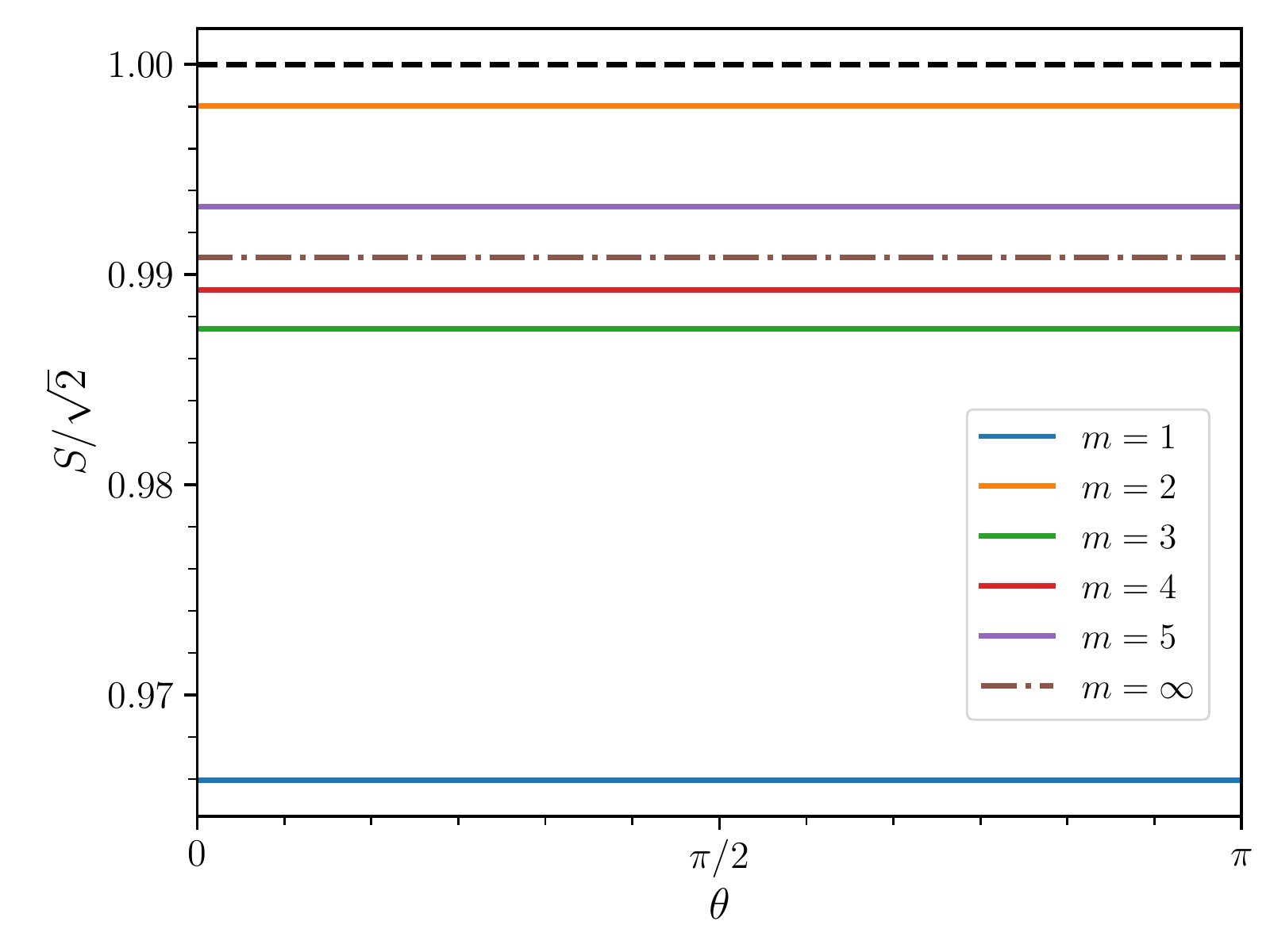} 
	\caption {Schmidt norm $S$ computed after evolution with the sequence ${\text{seq}}^*(2m+1)= [(H, F)^m, F]$ for $m=1,...,5$ as a function of the initial state parameter $\theta$ when $\phi=0$. The black dashed line indicates the maximum achievable value and the brown dashed-dotted line the asymptotic value for $m\to\infty$.}
	\label{fig1-new}
\end{figure}

We are interested in generating highly entangled states during a quantum walk independent of the initial state. Since the final amount of entanglement cannot be fully independent for all possible initial states \cite{Carneiro2005,Salimi2012}, we restrict the initial state to the class of localized states with zero relative phase, $\phi=0$ (see Eq.~(\ref{initial_general})). Hence, the problem reduces to finding a sequence of coin operators in time that generates entanglement independent of the initial state parameter $\theta$. For this we propose a sequence given by ${\text{seq}}^*(2m+1)= [(H, F)^m, F],\ m\in \mathbb{Z}$ for a quantum walk with $2m+1$ time steps. This sequence consists of an alternating application of the Hadamard and Fourier coin with an additional Fourier coin applied at the final time step and hence always describes a quantum walk with odd number of steps. In Fig.~\ref{fig1-new} we plot the Schmidt norm at the end of the quantum walk evolution with the proposed sequence for several different time steps as a function of the parameter $\theta$. One can easily see that the value of entanglement is always very close to the maximal amount possible and indeed independent of $\theta$ for each sequence. 
However it depends on the number of steps taken for short sequences, but quickly converges to a value close to $S/\sqrt{2}=0.99$ for larger values of $n$ (see Fig.~\ref{asympotic_HF}). The derivation of the asymptotic limit of this sequence is shown in Appendix \ref{Sec:AppendixA}. Each point in Fig.~\ref{asympotic_HF} is obtained after averaging over 1000 random angles $\theta$ and the zero variances confirm that the Schmidt norm is independent of the parameter $\theta$. Therefore, from now on we will refer to the sequence ${\text{seq}}^*$ as a universal entangler for the class of initial states defined by $\phi=0$.

In the following we will give an intuitive explanation of how the universal behavior emerges from this sequence. Generally, the Schmidt norm can be calculated from the reduced density matrix of the coin degree of freedom after tracing out the walker states. Representing the reduced density matrix $\rho$ as a vector on the Bloch sphere
\begin{equation} \label{reduced_dmatrix}
\rho =  \dfrac{1}{2}  {\it{I}} + \vec{\alpha} \vec{\sigma} ,
\end{equation}
where $\vec{\alpha}$ is the Bloch vector and 
$\vec{\sigma}$ is a vector of Pauli matrices, the Schmidt norm can be expressed in the form
\begin{equation} \label{Snorm-rdm}
S = \sqrt{ \dfrac{1}{2} + \mid {\vec{\alpha}}} \mid + \sqrt{ \dfrac{1}{2} - \mid {\vec{\alpha}}} \mid,
\end{equation}
which only depends on the norm of the Bloch vector $\vec{\alpha}$. During the evolution with the universal entangling sequence, the behavior of the Bloch vector $\vec{\alpha}$ follows a periodic pattern. Specifically, after each application of the Hadamard operator the Bloch vector points along the x-axis, while the subsequent application of the Fourier operator projects it onto the y-axis. For example, the sequence $[H, F, H]$ gives rise to $\vec{\alpha_3}=((\cos{\theta}+\sin{\theta)/4}, 0, 0)$, whereas after the sequence $[H, F, H, F]$ we obtain $\vec{\alpha_4}=(0, (-\cos{\theta}+4\sin{\theta)/16}, 0)$. At the end of the time evolution, an additional Fourier coin is applied, which rotates the Bloch vector into a $\theta$-dependent direction in the x-y plane with a norm that is independent of $\theta$. For example, after the evolution with the sequence ${\text{seq}}^*(5) = [H, F, H, F, F]$,  the Bloch vector calculates to $\vec{\alpha_5}=(\cos{\theta/16}, \sin{\theta/16}, 0)$ with $\mid{\vec{\alpha_5}} \mid=1/16$ and the Schmidt norm is independent of $\theta$ and approximately equal to 1.4114. The same property is also observed in the asymptotic limit when $m\to \infty$ (see Appendix \ref{Sec:AppendixA}).

\begin{figure}[t!]
	\includegraphics[scale=0.52, left]{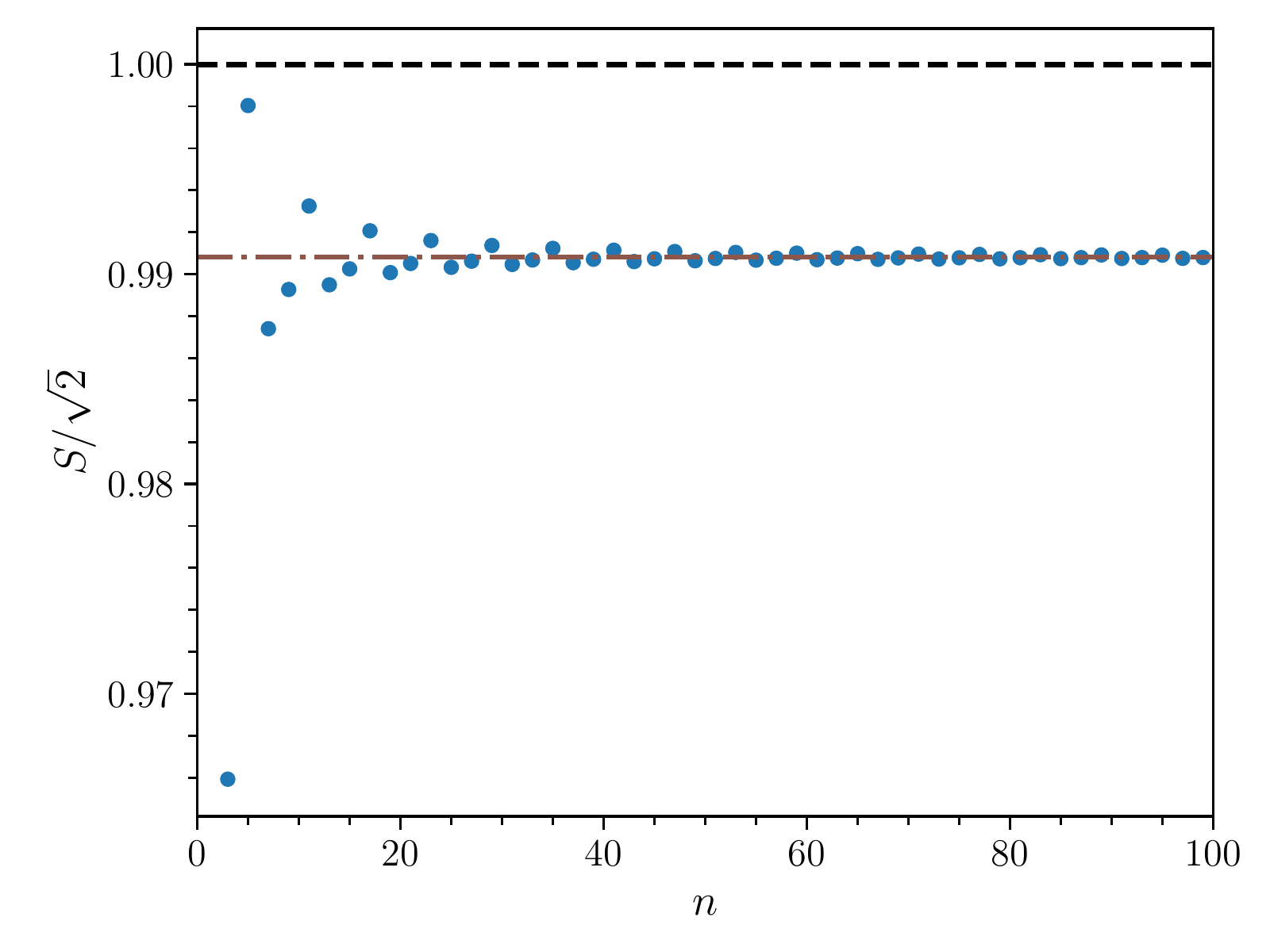} 
	\caption {Schmidt norm $S$ after evolution with the sequence ${\text{seq}}^*(2m+1)= [(H, F)^m, F]$ as a function of the number of steps $n = 2m+1$ (only odd time steps are displayed). Each point is an average over 1000 random initial states with $\phi=0$. The variances calculate to zero suggesting that the sequences ${\text{seq}}^*$ generate states with an amount of entanglement being independent of $\theta$. The dashed line denotes again the maximum achievable Schmidt norm  while the brown dashed-dotted line indicates the asymptotic value reached for $n=(2m+1)\to\infty$.}
	\label{asympotic_HF}
\end{figure}

\begin{figure}
	\includegraphics[scale=0.52, left]{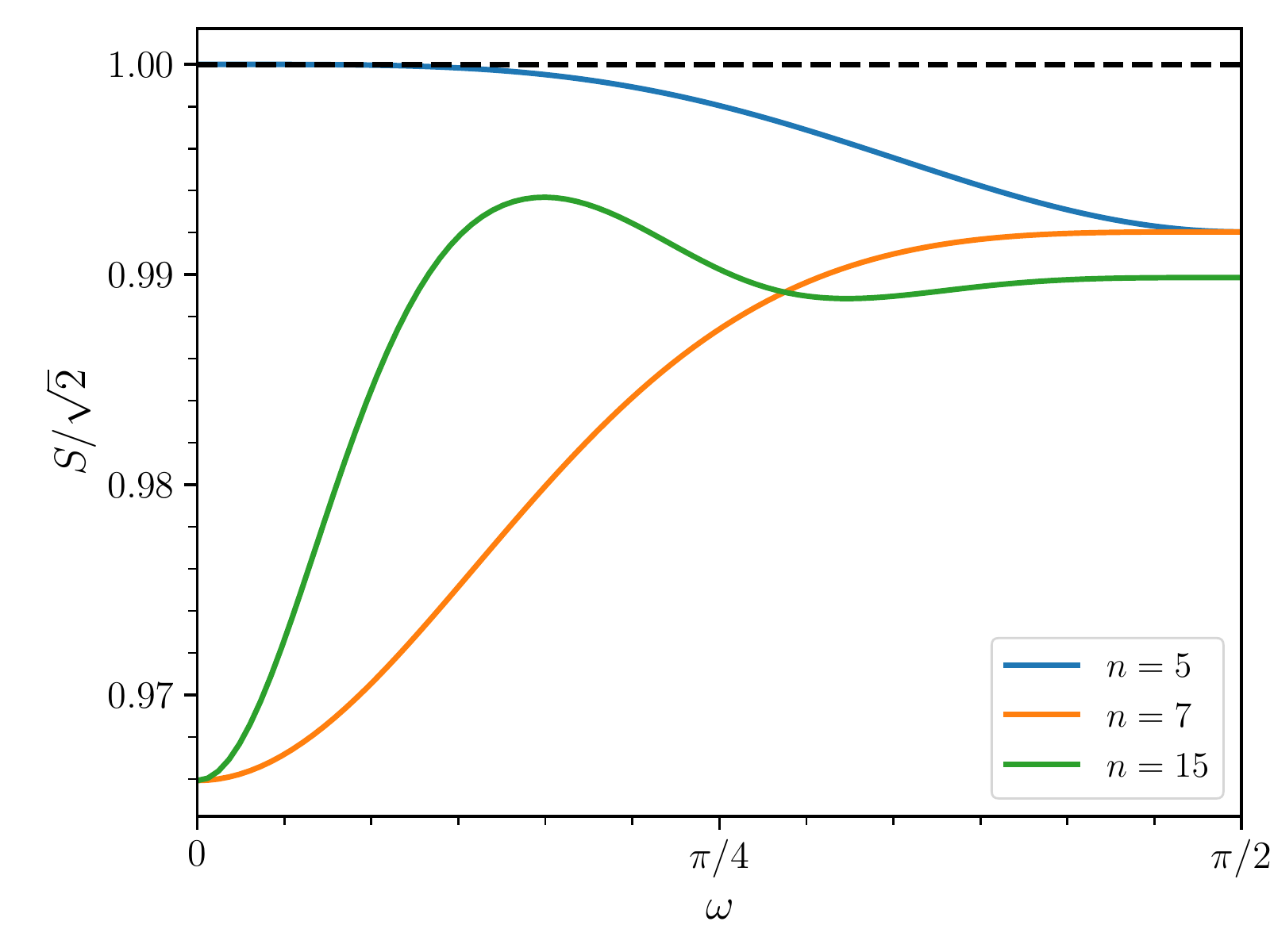} 
	\caption {Value of the Schmidt norm as a function of the generalized Hadamard operator parameter $\omega$ after the sequence $[(\tilde{H}(\omega), F)^m, F]$ for a 5-step (blue), 7-step (orange) and 15-step (green) quantum walk. Each point is an average over 1000 random initial states with $\phi=0$. The variances calculate to zero and the dashed line indicates the maximum achievable Schmidt norm.}
	\label{5steps_angle-omega}
\end{figure}

In order to better understand the behavior of the universal entangling sequence, we explore the role of the two coin operators $ H $ and $ F $. The Fourier operator seems to be of significant importance for generating highly entangled states. Generally, it increases the localization of the quantum state~\cite{Orthey2019} which has been associated with an enhancement in the entanglement~\cite{Chandrashekar2013}. On the other hand, the Hadamard operator belongs to the class of rotation matrices~\cite{Chandrashekar2007} and we have found that replacing it with a more general unbalanced operator does not change the universal behavior of the sequence. The generalized Hadamard operator $ \tilde{H} $ is given by
\begin{equation}
\tilde{H}(\omega) = \left[\begin{array}{rr}
\cos(\omega) & \sin(\omega) \\
\sin(\omega) & -\cos(\omega)
\end{array}\right],
\end{equation}
so that the sequence takes the new form of $[(\tilde{H}(\omega), F)^m, F]$. Fig.~\ref{5steps_angle-omega} shows the Schmidt norm after a 5, 7, and 15 step quantum walk as a function of the parameter $\omega$ for initial states with zero relative phase. 
Each data point was obtained after averaging over 1000 random angles $\theta$ of the initial state and the variance again calculates to zero in all cases. Therefore the amount of entanglement created is still independent of $\theta$. Moreover, the plot suggests that by properly choosing the parameter $\omega$ for a given length of the sequence, the performance of the universal entangling sequence can be improved and a state close to a maximally entangled state can be reached. 

Let us finally note that the effect of a nonzero relative phase $\phi$ in the initial state can be cancelled out in two ways using the phase operator Z given by
\begin{equation}
Z = \begin{bmatrix}
1 & 0 \\
0 & e^{-i\phi}
\end{bmatrix} .
\end{equation}
The phase operator can be applied either directly to the initial state or to the coin operators. In the latter case, the $H$ and $F$ operators are altered to $HZ$ and $FZ$, respectively. However, this requires that the relative phase of the initial state is known beforehand, which can be the case if the creation process of the initial state is deterministic.

\subsection{Optimal coin sequences}

Let us next address the question whether we can find coin sequences that perform better on average than the universal entangling sequence, i.e.~that generate higher values of entanglement across all initial states. To solve this optimization problem efficiently we employ the Q-Learning algorithm described in section \ref{Sec:RL}. We should emphasize that for a given number of steps $n$, the goal is to find the optimal sequence of coins out of the $2^n$ possible sequences that maximizes the Schmidt norm (the reward) for all initial states. Our RL framework allows us to solve for this objective due to the agents ignorance of the quantum state. Even though different initial quantum states are used for each episode, the agent has access only to the states defined by the history of actions and hence no information about the quantum state is used for training.

For a better comparison to the previous section, we again restrict the initial states to a subspace defined by $\phi=0$. For each episode of training, the remaining initial state parameter $\theta$ is sampled from a uniform distribution such that each episode is initialized with a different quantum state. The details of the training and the hyperparameters used can be found in Appendix \ref{Sec:AppendixB}.

\begin{figure}[t!]
	\centering
	\includegraphics[scale=0.52, left]{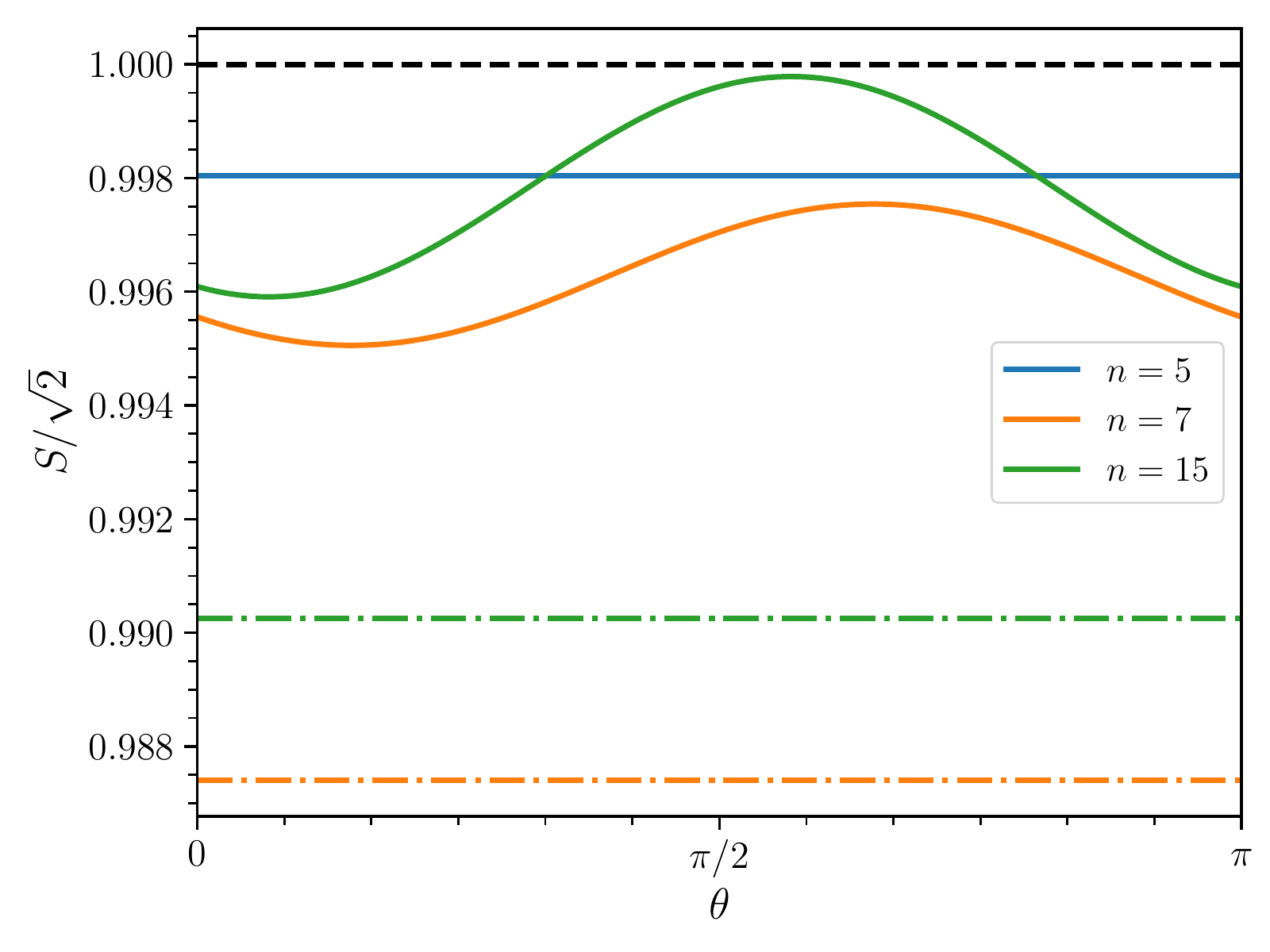} 
	\caption {Schmidt norm reached after an evolution of 5-step (blue), 7-step (orange), and 15-step (green) quantum walk with the optimal sequence (solid line) and the universal entangling sequence (dotted dashed line). The black dashed line denotes the maximum achievable Schmidt norm. In the case of a 5-step quantum walk the optimal and universal entangling sequence coincide. The optimal sequences are $[H,F,H,F,F]$, $[F,H,H,H,F,H,H]$, and $[F, H^7, F, H^6]$ respectively and were obtained using the Q-Learning algorithm.}
	\label{training_5steps}
\end{figure}

As an example we show the results of the RL optimization obtained for a 5, 7, and 15 step quantum walk in Fig.~\ref{training_5steps}. The Schmidt norm achieved by the optimal sequence is plotted as a function of the parameter $\theta$. Dashed lines of the same color correspond to the respective universal entangling sequence from the last section. Notice that in the case of a 5 step quantum walk the universal sequence and the optimal sequence coincide, i.e.~the RL agent finds $[H,F,H,F,F]$ to be optimal. For the cases of a 7 and 15 step quantum walk the optimal sequences differ from the universal ones and the obtained Schmidt norm is not independent of the initial state anymore. However, in both cases the amount of entanglement exceeds that of the universal sequence for all initial state parameters $\theta$.

In order to validate the result, we compared the reinforcement learning algorithm with a simple brute-force method for the case of the 5 step quantum walk. The brute-force algorithm explores all of the possible $ 2^5= 32 $ coin sequences for 1000 random initial states and computes the average Schmidt norm for each sequence. We find that the policy giving rise to the highest average entanglement is indeed the sequence the RL algorithm suggested previously: $ [H,F,H,F,F]$. While for quantum walks with only a few steps a simple brute-force method as described above is able to identify optimal policies, the RL algorithm becomes advantageous for larger numbers of time steps. The number of possible coin sequences grows exponentially with the number of steps and hence quickly becomes intractable by any brute-force method.

\begin{figure}[t!]
	\centering
	\includegraphics[scale=0.52, left]{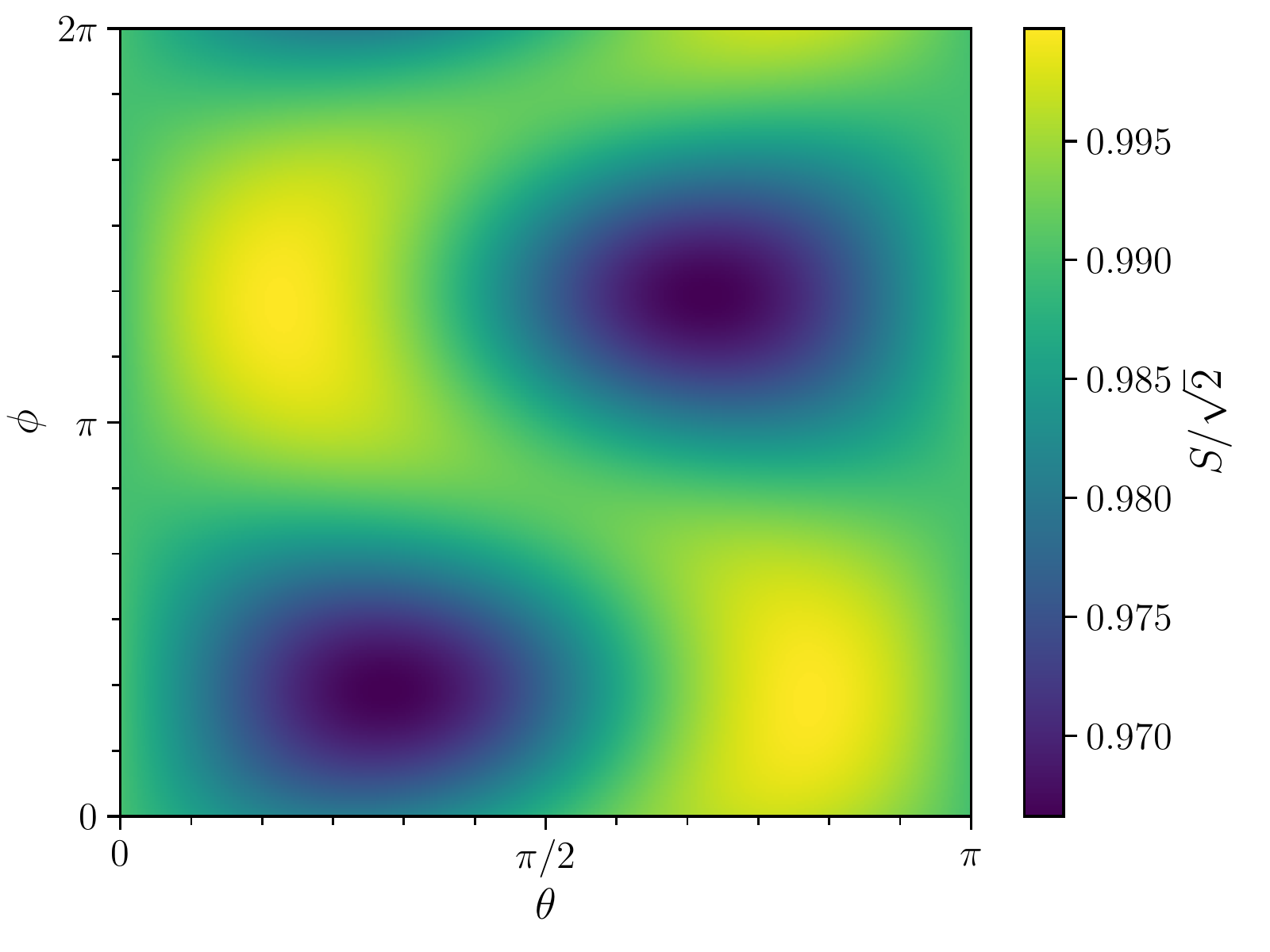}
	\caption {Schmidt norm obtained by evolving with the optimal policy $[F, F, H, H, H]$ as a function of the initial state parameters $\phi$ and $\theta$.}
	\label{ent_over_angles_random_phi}
\end{figure}

Finally, we train an RL agent on completely random initial states, where both $\phi$ and $\theta$ are uniformly sampled at the beginning of each episode. For a five step walk the optimal sequence suggested by the RL agent is $[F, F, H, H, H]$ and in Fig.~\ref{ent_over_angles_random_phi} we show the values of the achieved Schmidt norm as a function of the initial state parameters. One can see that the final amount of entanglement depends slightly stronger on the initial state compared to the previous cases where we only considered initial states with $\phi = 0$. This is not surprising since it is known that quantum walks of only a few steps cannot generate highly entangled states in a fully universal way for all initial states at the same time \cite{Carneiro2005,Salimi2012}. However, the RL algorithm is still able to identify a sequence that, at least on average, performs better than others.

\section{Conclusion}%
\label{Sec:Conclusions}%
We have discussed two different approaches for generating hybrid entanglement in a quantum walk. We first presented and studied an  entangling sequence consisting of a deterministic string of Hadamard and Fourier coin operators that created a universal amount of entanglement for all initial states with zero relative phase. Since this sequence works for any number of steps larger than two, it is valuable for experimental setting where the number of possible steps is limited. The second method was based on direct optimization of the coin sequence using reinforcement learning (RL), which is a technique that allows to also determine longer sequences where brute force optimisation is not possible. We have shown that this method allows to find coin sequences that yield high average values of entanglement over many initial states, which is particularly useful in cases where the initial state is not fully known, very noisy, or simply whenever it is required to generate highly entangled states independent of the initial state.

Our work therefore extends existing results that either achieve state independent entanglement only in the asymptotic limit of an infinite quantum walk \cite{Rigolin2013} or that can achieve maximal entanglement in a short sequence, but not independently of the initial state \cite{Innocenti2017, Gratsea2019}. Furthermore, the RL scheme we have presented can be useful in a variety of other, experimentally relevant settings. For example, the class of initial states that is optimized over can be restricted to match the experimental problem, such as a fixed initial state with noise. The RL objective can also be altered in different ways. One could for example choose to maximize the fidelity between the final state and a given target state. Another option is to apply the techniques to higher dimensional quantum walks \cite{Mackay2002}, quantum walks on graphs \cite{Aharonov2001}, or quantum walks involving more than one particle \cite{Omar2006}. Additionally, one could move to the continuous case and use deep reinforcement learning to directly optimize the parameters in the coin operator. 

Let us finally note that the universal entangling and optimized sequences give rise to qualitatively different probability distributions of the walker. The universal sequence produces a delocalised distribution whereas the optimised sequences  generate more localised one. This effect  will be a topic of future research.

\acknowledgements

This work was supported by OIST Graduate University and we are grateful for the help and support provided by the Scientific Computing section at OIST. We would also like to thank Alexander Dauphin for his helpful comments on the manuscript. A.G. acknowledges financial support from the Spanish Ministry MINECO (National Plan 15 Grant: FISICATEAMO No. FIS2016-79508-P, FPI), European Social Fund, Fundació Cellex, Generalitat de Catalunya (AGAUR Grant No. 2017 SGR 1341, CERCA/Program), ERC AdG NOQIA, EU FEDER, MINECO-EU QUANTERA MAQS, the National Science Centre, Poland-Symfonia Grant No. 2016/20/W/ST4/00314 and the Marie Sklowdowska-Curie-COFUND action under the European Union’s Horizon
2020 research and innovation programme (GA 847517).


\bibliographystyle{apsrev4-1}
\bibliography{manuscript.bib}

\appendix

\section{Asymptotic limit of the universal entangling coin sequence}
\label{Sec:AppendixA}%
In this Appendix we derive the asymptotic limit of the coin reduced density matrix under the universal entangling sequence, i.e. ${\text{seq}}^*(2m+1)= [(H, F)^m, F]$ with $m\rightarrow\infty$, which allows us to calculate the asymptotic value of the Schmidt norm and prove its independence of the initial state angle $\theta$. We follow the approach of Ref.~\cite{Hin2014,Brun2003} where the evolution of the reduced density matrix of the coin degree of freedom is directly computed through an effective superoperator in Fourier space.

The quantum walk shift operator $S$ of Eq.~\eqref{uniS} can be expressed in momentum space after performing a Fourier transform defined by $|k\rangle=\sum_{x} e^{i k x}|x\rangle$, which leads to
\begin{equation}
    S_k = \ket{k}\bra{k}\otimes \left(e^{-ik}\ket{\downarrow}\bra{\uparrow} + e^{ik} \ket{\uparrow}\bra{\downarrow}\right).
\end{equation}
The combined effect of the shift and coin operator can therefore be reduced to a $2\times 2$ matrix acting on the coin degree of freedom only and for the Hadamard and Fourier coin we obtain
\begin{align}
\label{Eq:H}
S_k{H}=&\mathbb{1}_k \otimes\frac{1}{\sqrt{2}}\left(\begin{array}{cc}
e^{i k} & -e^{i k} \\
e^{-i k} & e^{-i k}
\end{array}\right),\\
\label{Eq:F}
S_k{F}=&\mathbb{1}_k \otimes\frac{1}{\sqrt{2}}\left(\begin{array}{cc}
ie^{i k} & e^{i k} \\
e^{-i k} & ie^{-i k}
\end{array}\right).
\end{align}
These operators act on the full quantum state $\ket{\psi}$ (coin and momentum degree of freedom), however, we can also directly work in the reduced space of the coin which can be represented as a vector on the Bloch sphere as
\begin{equation}
\rho=\text{Tr}_k\left(\left|\psi\right\rangle\left\langle\psi\right|\right) =\alpha_{0} I+\alpha_{1} \sigma_{1}+\alpha_{2} \sigma_{2}+\alpha_{3} \sigma_{3}.
\end{equation}
For an arbitrary initial state of Eq.~\eqref{initial_general} the Bloch vector components yield
\begin{equation}
\label{Eq:rho}
\vec{\rho}_0=\left(\begin{array}{c}
\alpha_{0} \\
\alpha_{1} \\
\alpha_{2} \\
\alpha_{3}
\end{array}\right)=\frac{1}{2}\left(\begin{array}{c}
1 \\
\cos \varphi \sin \theta \\
-\sin \varphi \sin \theta \\
\cos \theta
\end{array}\right).
\end{equation}
During the quantum walk evolution the reduced density matrix transforms according to an effective evolution superoperator $L_k$ and after $n$ steps of the quantum walk is given by
\begin{equation}
\rho_{n}=\int_{-\pi}^{\pi} \frac{d k}{2 \pi} \left(L_{k}\right)^{n}\rho_0.
\end{equation}
Using the vector notation of Eq.~\eqref{Eq:rho}, the operator $L_k$ can be represented as a $4\times 4$ matrix. The matrix entries are obtained after working out how each of the Pauli matrices transforms under the combined effect of shift and coin operator, i.e. Eqs.~\eqref{Eq:H} and \eqref{Eq:F}.
For the case of the Hadamard and Fourier coin the superoperators compute to
\begin{align}
L_{k}^{H}&=\left(\begin{array}{cccc}
1 & 0 & 0 & 0 \\
0 & 0 & \sin 2 k & \cos 2 k \\
0 & 0 & \cos 2 k & -\sin 2 k \\
0 & -1 & 0 & 0
\end{array}\right),\\
L_{k}^{F}&=\left(\begin{array}{cccc}
1 & 0 & 0 & 0 \\
0 & \cos 2 k & 0 & -\sin 2 k \\
0 & -\sin 2 k & 0 & -\cos 2 k \\
0 & 0 & 1 & 0
\end{array}\right).
\end{align}
Hence, two steps of the quantum walk with a Hadamard coin applied at the first time step and a Fourier coin applied at the second time step, give rise to the following superoperator
\begin{align}
L_{k}^{HF}&=L_{k}^{F}L_{k}^{H}\nonumber\\
&=\left(\begin{array}{cccc}
1 & 0 & 0 & 0 \\
0 & \sin 2 k & \sin 2 k \cos 2 k& \cos^2 2 k \\
0 & \cos 2 k & -\sin^2 2 k & -\sin 2 k \cos 2 k \\
0 & 0 & \cos 2 k & -\sin 2 k
\end{array}\right).
\end{align}
Since we are interested in the long time behavior, we first diagonalize the matrix above before exponentiating it to the desired power. The eigenvalues are given by
\begin{equation}
\lambda_{0}=1,\ \lambda_{1}=1,\ \lambda_{2}=e^{i(\gamma+\pi)},\ \lambda_{3}=e^{-i(\gamma+\pi)},
\end{equation}
with
\begin{equation}
\cos \gamma=\frac{1}{2}(1+\sin^2 2 k).
\end{equation}
After $n=2m$ steps of the quantum walk with a Hadamard and Fourier coin applied at alternating time steps we obtain
\begin{equation}
\left(L_{k}^{HF}\right)^{m}=B\left(\begin{array}{cccc}
1 & 0 & 0 & 0 \\
0 & 1 & 0 & 0 \\
0 & 0 & e^{i m(\gamma+\pi)} & 0 \\
0 & 0 & 0 & e^{-i m(\gamma+\pi)}
\end{array}\right) B^{\dagger},
\end{equation}
where the matrix $B$ contains the corresponding eigenvectors as column entries
\begin{equation}
B=\left(\begin{array}{cccc}
1 & 0 & 0 & 0 \\
0 & v_{11} & v_{12} & v_{13} \\
0 & v_{21} & v_{22} & v_{23} \\
0 & v_{31} & v_{32} & v_{33}
\end{array}\right).
\end{equation}
When taking the limit $m\rightarrow\infty$, the oscillatory terms $e^{\pm i m(\gamma+\pi)}$ vanish due to the stationary phase theorem. Therefore, we get the following expression for the asymptotic superoperator
\begin{equation}
\left(L_{k}^{HF}\right)^{m} \xrightarrow[m \to \infty]{}
\left(\begin{array}{cccc}
1 & 0 & 0 & 0 \\
0 & \left|v_{11}\right|^{2} & v_{11} v_{21}^{*} & v_{11} v_{31}^{*} \\
0 & v_{21} v_{11}^{*} & \left|v_{21}\right|^{2} & v_{21} v_{31}^{*} \\
0 & v_{31} v_{11}^{*} & v_{31} v_{21}^{*} & \left|v_{31}\right|^{2}
\end{array}\right),
\end{equation}
which only involves the components of the first eigenvector
\begin{equation}
\begin{aligned}
\overrightarrow{v_{1}} &=\left(\begin{array}{l}
v_{11} \\
v_{21} \\
v_{31}
\end{array}\right) =\frac{ \cos 2k}{\sqrt{4-(\sin^2 2 k +1)^2}}\left(\begin{array}{c}
1+ \sin 2 k\\
\cos 2 k \\
1 - \sin 2 k
\end{array}\right).
\end{aligned}
\end{equation}
The superoperator $L_{k}^{*}$ of the universal entangling sequence is obtained by acting with an additional final Fourier superoperator $L_k^F$
\begin{equation}
L_{k}^{*}(m) = L_{k}^{F}\left(L_{k}^{HF}\right)^{m}.
\end{equation}
The asymptotic limit of the reduced density matrix can then be calculated by performing the momentum integrals for each matrix entry separately giving rise to
\begin{align}
{\vec{\rho}}_{\infty}^* &=\lim _{m \rightarrow \infty} \int_{-\pi}^{\pi} \frac{d k}{2 \pi} L_{k}^{*}(m)\vec{\rho}_0\nonumber\\
&=
\left(\begin{array}{cccc}
1 & 0 & 0 & 0 \\
0 & 0 & -1 + \dfrac{2}{\sqrt{3}} & 2-\sqrt{3} \\
0 & -2 + \sqrt{3} & 1 - \dfrac{2}{\sqrt{3}} & 0 \\
0 & 0 & -1 + \dfrac{2}{\sqrt{3}} & 0
\end{array}\right)\left(\begin{array}{c}
\alpha_{0} \\
\alpha_{1} \\
\alpha_{2} \\
\alpha_{3}
\end{array}\right)\nonumber\\
&=
\frac{1}{2}\left(\begin{array}{c}
1 \\
\left(2-\sqrt{3}\right) \cos \theta \\
\left(-2+\sqrt{3}\right) \sin \theta  \\
0
\end{array}\right).
\end{align}
In the last line we used that $\phi=0$ for the initial states considered here. The final state lies in the $x-y$ plane of the Bloch sphere with a norm independent of the angle $\theta$. As a consequence the Schmidt norm defined in Eq.~\eqref{Snorm-rdm}, which is only a function of the length of the Bloch vector, is also independent of $\theta$ and computes to
\begin{align}
    S &= \sqrt{ \dfrac{1}{2} + \dfrac{1}{2}(2-\sqrt{3})} + \sqrt{ \dfrac{1}{2} - \dfrac{1}{2}(2-\sqrt{3})}\nonumber\\
    &\sim 0.9908\times \sqrt{2}.
\end{align}
This value matches the asymptotic behavior we observe in  Fig.~\ref{asympotic_HF} of the main text.

\section{Details of the RL training procedure} \label{Sec:AppendixB}%
\begin{figure*}[h]
	\begin{subfigure}
	\centering
		\includegraphics[width=0.45\linewidth]{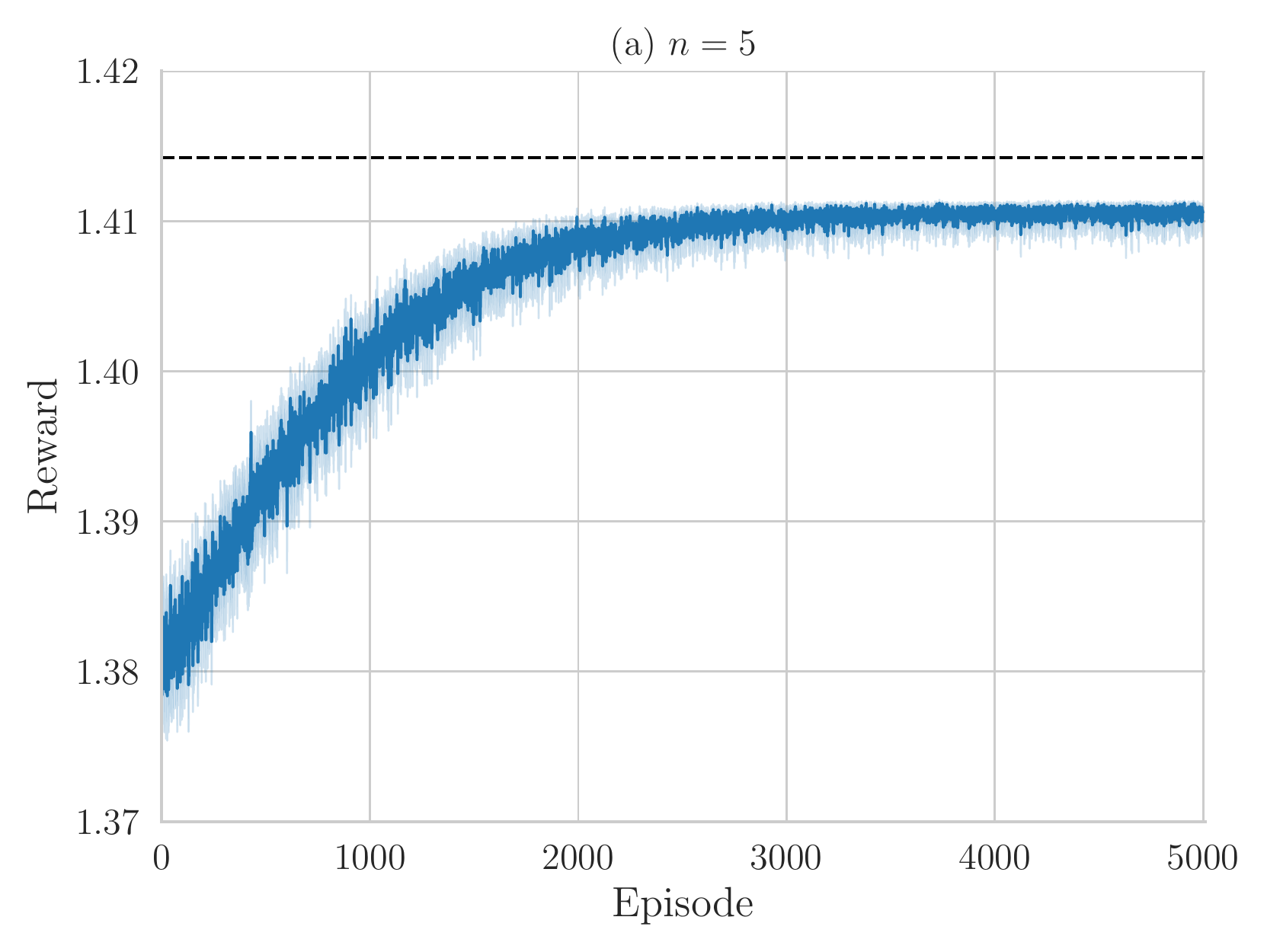}
	\end{subfigure}
	\begin{subfigure}
		\centering
		\includegraphics[width=0.45\linewidth]{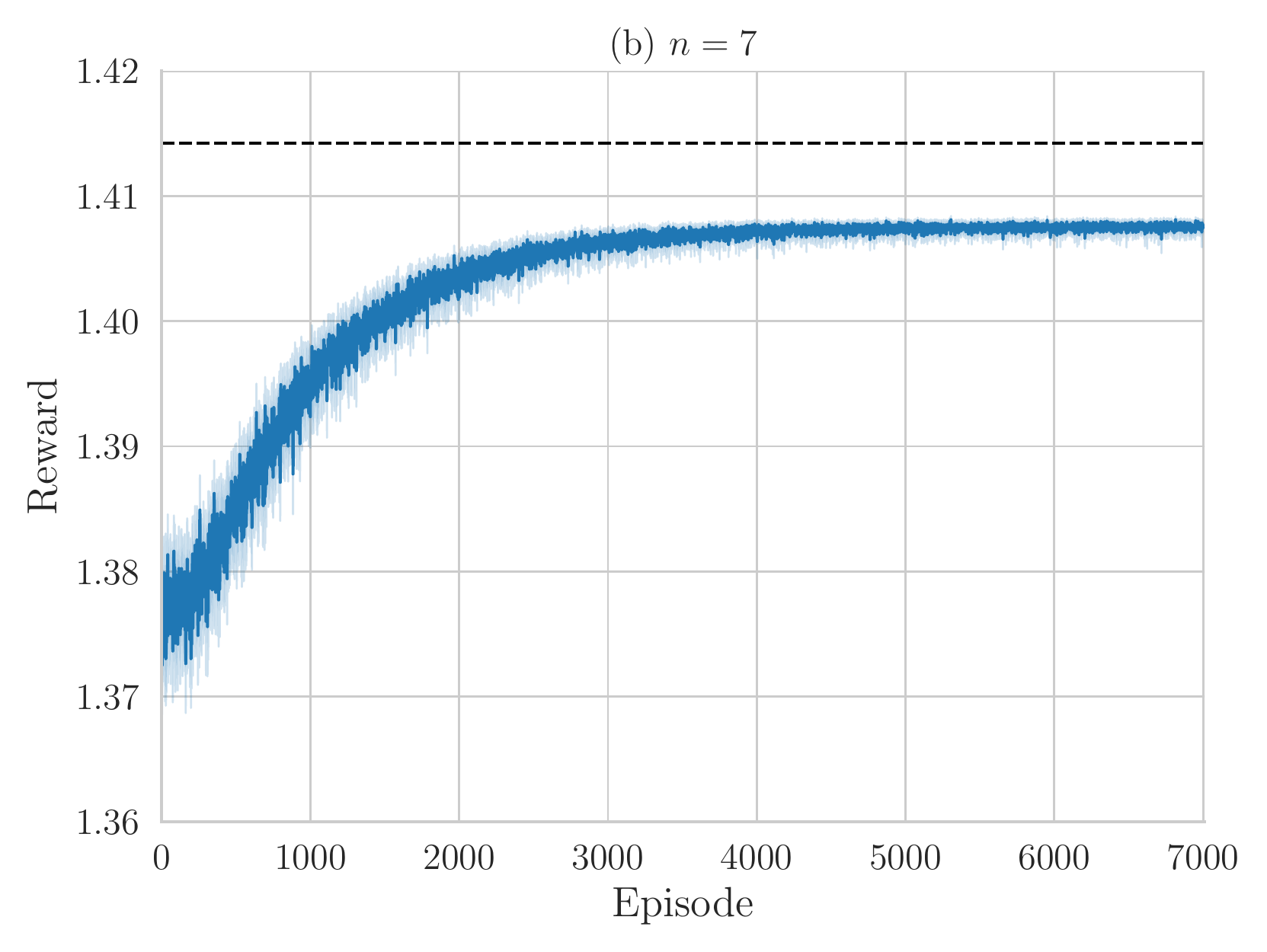}
	\end{subfigure}
	\newline
	\begin{subfigure}
		\centering
		\includegraphics[width=0.45\linewidth]{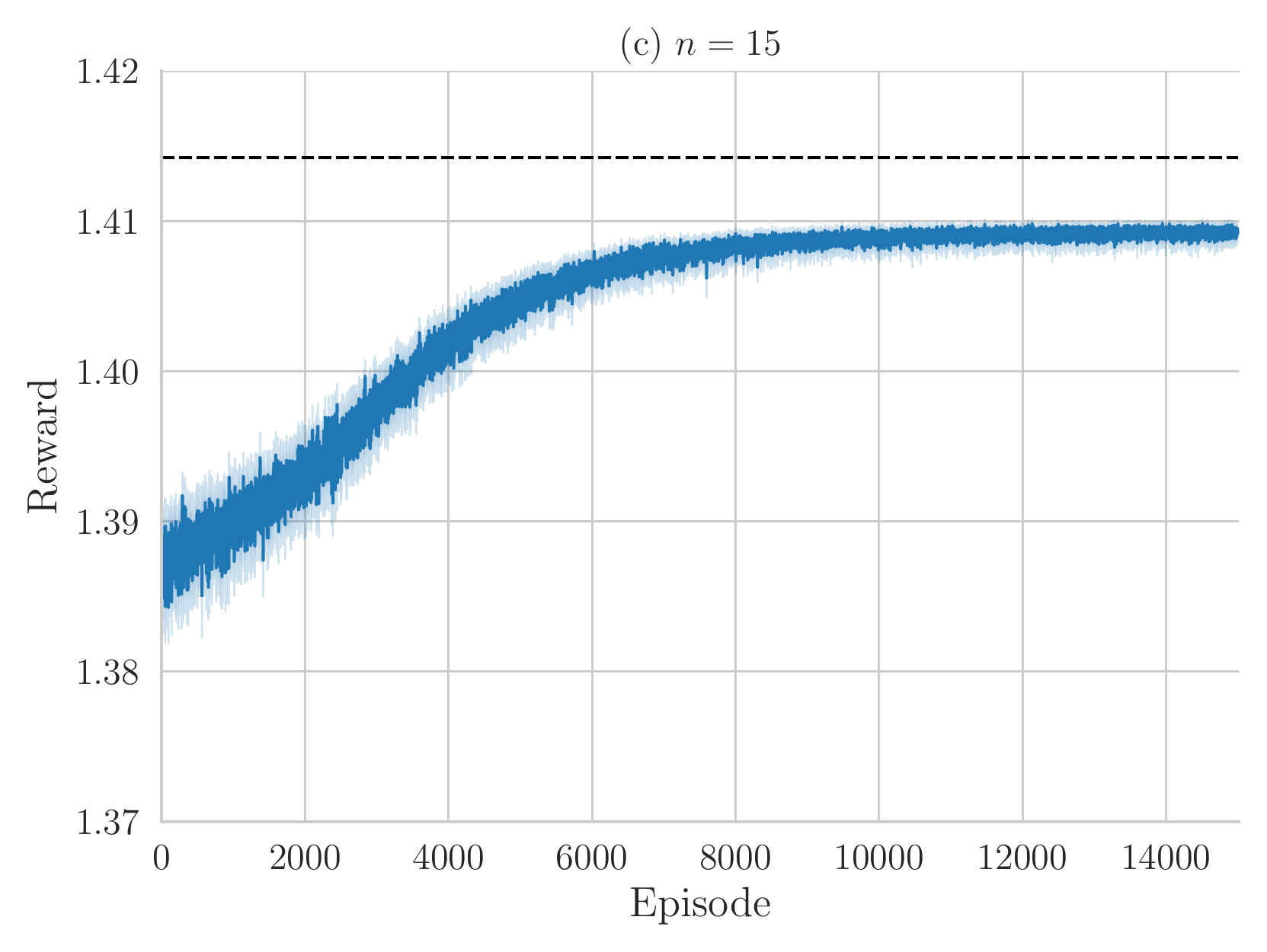}
	\end{subfigure}
	\begin{subfigure}
		\centering
		\includegraphics[width=0.45\linewidth]{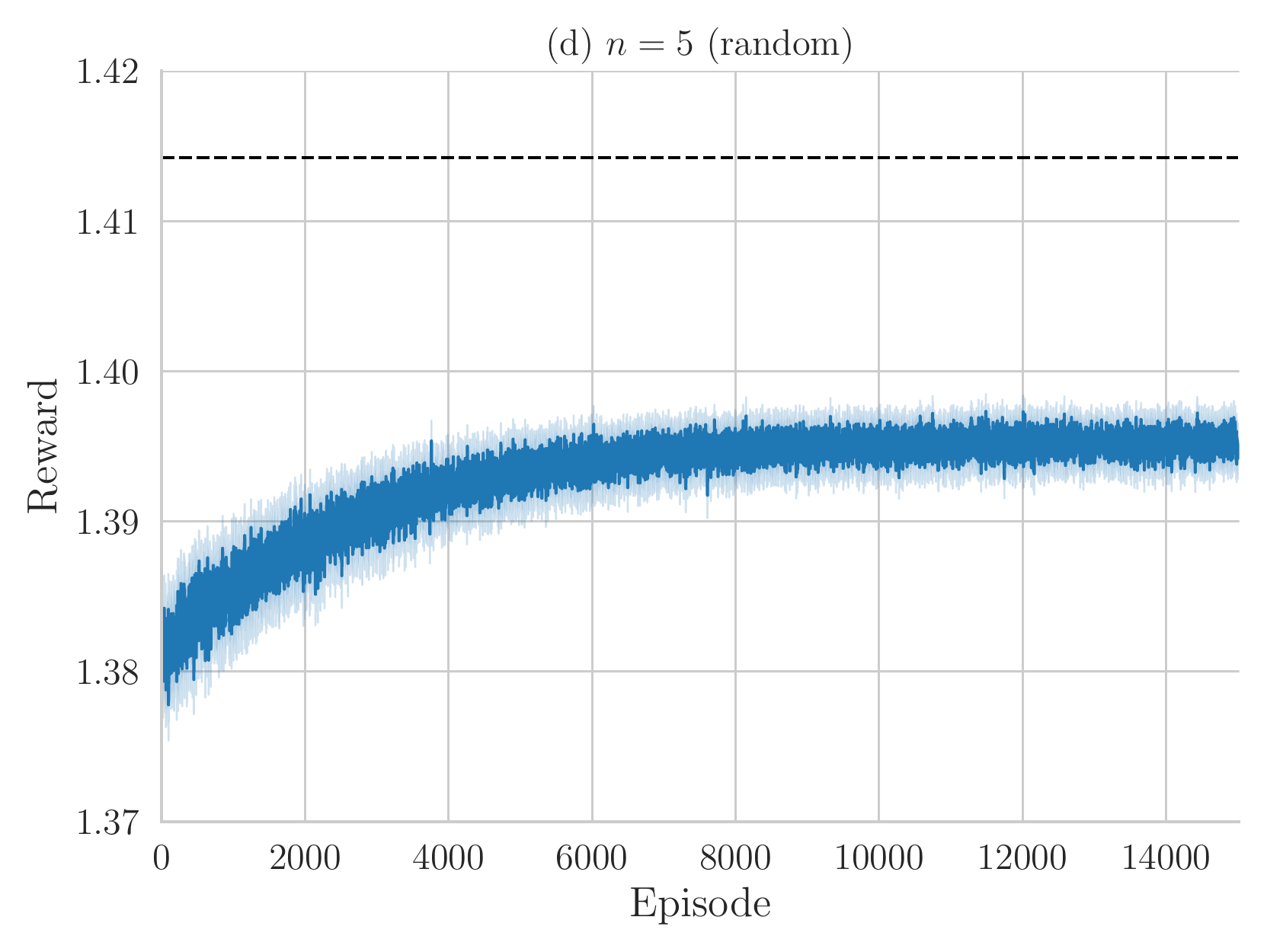}
	\end{subfigure}
	\caption{Learning curves for the optimization problems discussed in the main text. The episodic reward (Schmidt norm) is averaged over 300 ((a), (b)) or 400 ((c), (d)) independent runs. The light blue area corresponds to the confidence interval and dashed lines denote the maximally achievable reward of $\sqrt{2}$. (a)-(c) Learning curves for the 5, 7, and 15 step quantum walk where the initial state parameter $\phi$ is set to zero and the parameter $\theta$ is sampled from a uniform distribution at the beginning of each new episode. (d) Learning curve for the 5 step quantum walk where both initial state parameters $\phi$ and $\theta$ are sampled at the beginning of each training episode.}
	\label{fig6}
\end{figure*}

	All instances of training were performed using the Q-Learning algorithm \cite{Watkins1992} with Q values initialized to zero. We found a learning rate of $\alpha = 0.7$ to give the best results overall. The exploration parameter $\epsilon$ decays exponentially throughout the training from an initial value of $\epsilon_i=0.9$ to a final value of $\epsilon_f=0.01$. The only parameters that were changed for obtaining the different results in the main text are the number of episodes of training and the number of independent runs. In Fig.~\ref{fig6} on the next page we show the learning curves of the RL agent for a 5, 7 and 15 step quantum walk that are further discussed in the main text.

\clearpage
\onecolumngrid

\end{document}